%% file: main.tex
\documentclass[11pt,fleqn,amsmath,a4paper]{article}
\usepackage{graphicx} 
\usepackage{xspace}
\usepackage{placeins} 
\usepackage{xcolor}
\usepackage{amsmath}
\usepackage{amssymb}
\usepackage{rotating} 
\usepackage{booktabs}
\usepackage{authblk} 
\usepackage{tocbibind} 
\usepackage{hyperref} 
\usepackage[left=1in, right=1in]{geometry}
\hypersetup{
    colorlinks=true,
    linkcolor=blue,
    filecolor=magenta,      
    urlcolor=blue,
    pdftitle={ANUBIS EPPSU inputs},
    pdfpagemode=FullScreen,
    }
\usepackage[backend=biber,style=phys,eprint=true,hyperref=true,biblabel=brackets]{biblatex}
\usepackage{tcolorbox}
\usepackage{enumitem}
\newcommand{\eg}{\mbox{\itshape e.g.}\xspace}
\newcommand{\ie}{\mbox{\itshape i.e.}\xspace}
\newcommand{\etc}{\mbox{\itshape etc.}\xspace}

\input{style}
\title{2026 ESPPU input from the ANUBIS Collaboration}
\author{ANUBIS Collaboration}

\date{Last update: \today}

\begin{document}

\begin{titlepage}
\maketitle
\begin{abstract}
 It is imperative for us as a particle physics community to fully exploit the physics potential of the High-Luminosity LHC.
 This calls for us not to leave any stone unturned in the search for Beyond the Standard Model (BSM) physics.
 Many BSM models that address fundamental questions of physics like the particulate nature of dark matter, the matter-antimatter asymmetry in the Universe, small but non-zero neutrino masses \etc, predict Long-Lived Particles (LLPs) with macroscopic lifetimes of $\tau>10^{-10}$~s.
 The challenge in searching for BSM models with LLP signatures at the HL-LHC is that it requires the complementary interplay of general purpose detectors like ATLAS, CMS, and LHCb; dedicated detectors situated close to the beamline including the proposed Forward Physics Facility~(FPF); and dedicated detectors covering a large decay volume at a reasonable solid angle transverse to the beamline, \ie a Transverse Physics Facility~(TPF). 
 Hence, it is of vital importance to realise a TPF in order to expand dramatically the physics coverage within long-lived particle searches to harvest the physics at the HL-LHC fully. 
 A TPF may be composed of several experiments based at the HL-LHC.
 In this document, we propose that the community realise the ANUBIS experiment as part of a TPF.        
\end{abstract}

\end{titlepage}

\section{Scientific context}
Despite the many successes of experimental particle physics over the last few decades, such as the development of the Standard Model~(SM), measuring its parameters to high precision, and the discovery of the Higgs boson, many fundamental questions remain unanswered. This includes the exact nature of dark matter~\cite{Hall:2010jx,Cheung:2010gk,Zurek:2013wia}; a mechanism for the the matter-antimatter asymmetry of the Universe; and small but non-zero neutrino masses. Many Beyond the Standard Model (BSM) theories that can answer these questions predict new particles with macroscopic lifetimes of $\tau>10^{-10}$~s, referred to as Long-Lived Particles (LLPs). 
Another motivation for LLP searches is the absence of a significant observation of BSM particles at existing General Purpose Detectors (GPD) such as ATLAS~\cite{Aad:1129811}, CMS~\cite{Chatrchyan:1129810}, LHCb~\cite{Alves:1129809} \etc, which can be explained if the new BSM particles reside in a `Hidden-Sector' with only a weak coupling to SM particles. 
This weak coupling naturally leads to LLPs that act as a `portal-particle' or mediator between the SM and the Hidden Sector. 
As such LLPs are a generic feature of many BSM models, including supersymmetry~\cite{Barbier:2004ez,Giudice:1998bp}, hidden sector models~\cite{Han:2007ae,Strassler:2006im}, and exotic Higgs decays~\cite{Chacko:2005pe,Cai:2008au}. 
In general, $\taullp$  is a free parameter in BSM models, motivating a large variety of potential search signatures which require a broad experimental approach.\\

Although the GPD experiments at the LHC could explore a significant fraction of the LLP scenario space~\cite{Alimena:2019zri, CMS:2021sch, CMS:2021juv}, they are fundamentally limited by their finite detection volumes, intricate backgrounds (\eg pile-up), and trigger requirements designed for prompt particles. For example, a neutral LLP with sufficient lifetime could traverse a GPD and subsequently decay unregistered outside its active volume. 
The logical next step to extend the coverage to scenarios involving very long-lived particles with $c\tau >\mathcal{O}(10~\rm m)$ is to extend the decay volume. 
In recent years a set of dedicated LLP experiments has been proposed such as those in Refs.~\cite{Bauer:2019vqk,Chou:2016lxi,Curtin:2018mvb, MATHUSLA:2020uve,Gligorov:2018vkc,Gligorov:2017nwh} and some have been realised, \eg FASER~\cite{FASER:2018eoc}.\\

\begin{figure}[htpb]
    \centering
    \includegraphics[width=0.55\linewidth]{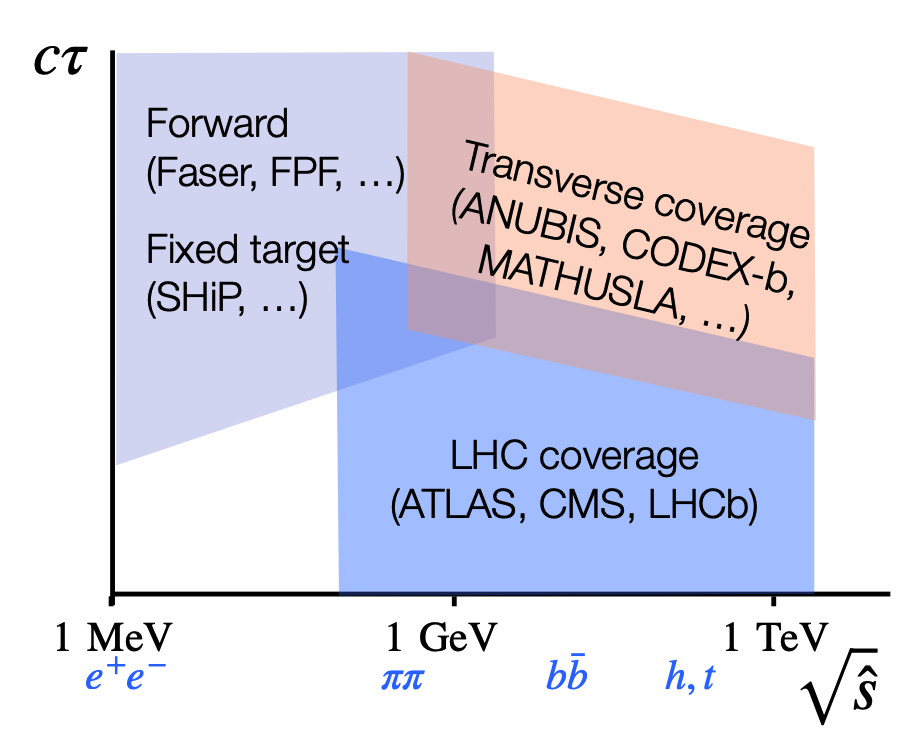}
    \caption{A schematic representation of the coverage of different types of experiments to LLPs as a function of the partonic centre-of-mass energy, $\sqrt{\hat{
    s}}$.}
    \label{fig:LLP_Complementarity}
\end{figure}

LLP searches at colliders can be grouped into three categories: searches using GPDs; Forward/Beam Dump experiments that lie longitudinally along beamlines; and Transverse experiments that are positioned transverse to the beamline. Each of these can probe both unique and complementary regions of the available parameter-space of LLP models, as shown in Figure~\ref{fig:LLP_Complementarity} and summarised below:
\vspace{-2mm}
\begin{itemize}[itemsep=0mm]
\item 
GPDs can cover the entire kinematic parameter space available at the LHC, but are constrained in their lifetime reach to $\tau \lesssim 10$~ns by their finite volume and suffer from large backgrounds from hadronic interactions and pile-up. 
\item 
The Forward/Beam Dump experiments are sensitive to longer lifetimes, but cannot probe either heavy LLPs with $\mllp>\mathcal{O}(10~\GeV)$ or LLPs produced by massive mediators at the electroweak scale and above, due to kinematic constraints. 
\item 
The Transverse experiments (and ANUBIS in particular) close the remaining sensitivity gap by probing the entire kinematic parameter space including mediator masses well above the electroweak scale at long lifetimes $\tau \gtrsim 10$~ns that cannot be tested elsewhere.
\end{itemize}

In order to fully exploit the physics potential of the HL-LHC, a varied LLP search programme including GPDs, a Forward Physics Facility~(FPF), and a Transverse Physics Facility~(TPF) must be established~\cite{Antel:2023hkf}. 
At future colliders, GPDs should be developed with LLPs in mind from the beginning, featuring both an FPF and a TPF. 
Given the uncertainty about the next generation collider and the timescale for obtaining data, it is imperative to realise both Forward and Transverse facilities on the shorter timeline of the HL-LHC. 

There are several forward experiments that are currently operational, \eg FASER~\cite{Feng:2017uoz}, as well as plans to build a dedicated experimental hall with several Forward experiments collectively known as the Forward Physics Facility~\cite{Feng:2022inv}. However, to date there have been no Transverse experiments constructed or operated. This is a considerable gap in the potential coverage, since LLPs around or above the electroweak scale with substantial lifetimes $\tau\gtrsim10$~ns can not be explored at a FPF nor at a GPD, calling for a TPF to be constructed.

There are currently several mature proposals for transverse LLP experiments that are based near different LHC experiments at CERN, which adopt slightly different detector technologies, background mitigation strategies, and geometries. These are: ANUBIS~\cite{Bauer:2019vqk}, which would be based in the ATLAS experimental cavern; CODEX-b~\cite{Gligorov:2017nwh}, which would be based near the LHCb experimental cavern; and MATHUSLA~\cite{Curtin:2018mvb}, which would be based on the surface above the CMS experiment. It is our strong recommendation that at least one of these should be built, with our preference for the ANUBIS experiment. However, if multiple are built then a multi-experiment, robust TPF would allow for much better exploration of the transverse parameter-space for LLPs.

\section{Scientific objectives for ANUBIS}
\begin{mybox}
\begin{itemize}[itemsep=0mm]
\item 
Fully exploit the physics potential of HL-LHC and the existing physics infrastructure.
\item 
Extend the sensitivity of the ATLAS detector and provide complementary sensitivity to Forward/Beam Dump experiments like FASER and SHiP.
\item 
Provide the first direct probe for long-lived particles with masses above 10~GeV produced at the electroweak scale with lifetimes up to $\mathcal{O}(1~\mu\text{s})$.
\item 
Provide constraints on 
\vspace{-2mm}
\begin{itemize}[itemsep=0mm]
\item 
non--trivial dark matter scenarios~\cite{Hall:2010jx,Cheung:2010gk,Zurek:2013wia};
\item 
baryogenesis--inspired models~\cite{Barry:2013nva};
\item 
Higgs portal models like Neutral Naturalness~\cite{Giudice:1998bp,Burdman:2006tz};
\item 
\etc
\end{itemize}
\end{itemize}
\end{mybox}

\section{Methodology}
The golden signature targeted by ANUBIS is a neutral LLP that decays within its air-filled active volume, which is then reconstructed as a displaced vertex from charged particle tracks. 
Highly efficient tracking detectors are required to reconstruct tracks and ultimately displaced vertices. 
For optimal physics performance, a per-layer hit efficiency of more than 98\% is required.
Also, to reject instrumental backgrounds from cosmic rays and beam-induced processes, a time resolution of $\lesssim0.5\text{ ns}$ is necessary. 
Finally, to accurately reconstruct displaced vertex positions, a spatial resolution of $\lesssim0.5\text{ cm}$ and an angular resolution of $\lesssim0.01$~rad is required~\cite{Bauer:2019vqk}.
The above detector performance benchmarks are also beneficial to probe other signatures, \eg, charged LLPs,  using ANUBIS' superb time-of-flight capabilities.

\begin{figure}[htpb]
    \centering
    \includegraphics[width=0.48\linewidth]{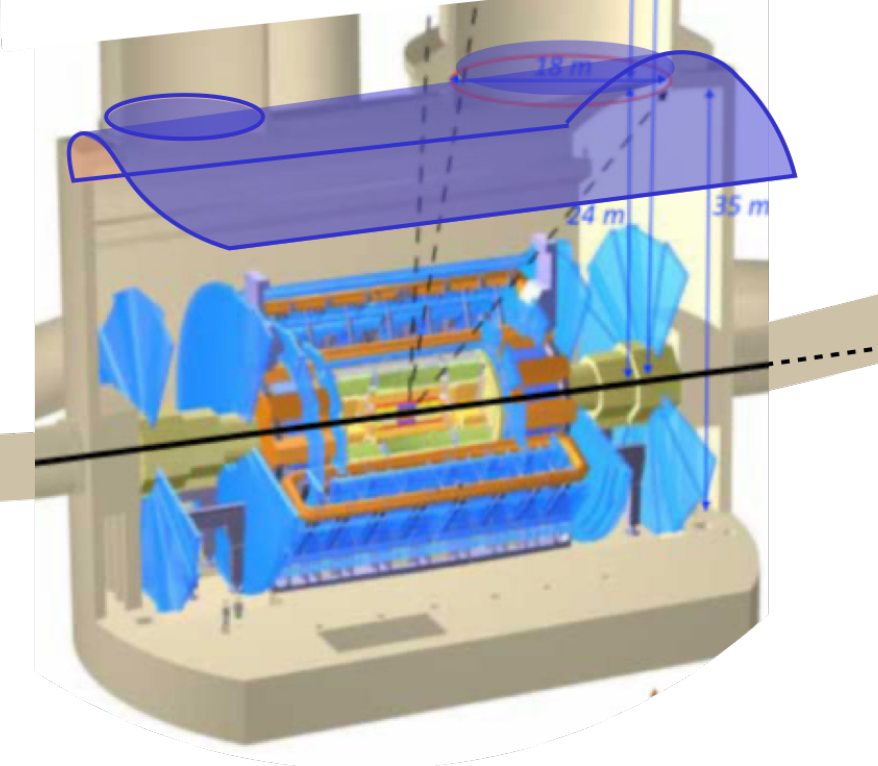}
    \includegraphics[width=0.44\linewidth]{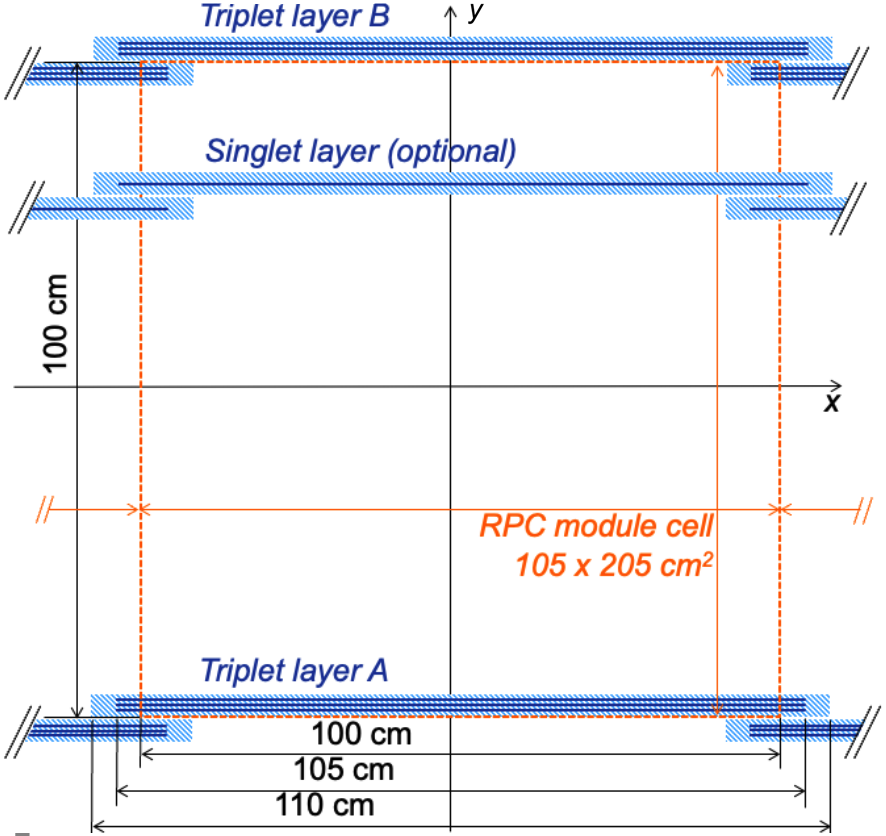}
    \caption{(left)~The proposed location for the ANUBIS detector~(blue) in the ATLAS experimental cavern, attached the the ceiling of the cavern dome. (right)~The layout of the RPC tracking stations within ANUBIS.}
    \label{fig:ANUBIS_Geometry}
\end{figure}

The core idea behind the ANUBIS detector is to use the space between the ATLAS detector and the ATLAS detector cavern ceiling as an air-filled active decay volume. 
ANUBIS will be composed of multiple layers of tracking detectors, which are installed directly onto the ceiling of the ATLAS cavern or at the bottom of the PX14 and PX16 service shafts to the cavern, see Figure~\ref{fig:ANUBIS_Geometry}~(left). This is done to maximise the decay volume ANUBIS is sensitive to, without the requirement for additional large-scale civil engineering. 
ANUBIS would then be situated at a radius of $\mathcal{O}(20-30)\text{ m}$ from the Interaction Point (IP), providing an unprecedentedly large decay volume for LLP searches using an instrumented area of $1,600~\metre^2$.
The detector performance requirements outlined above can be fulfilled using ATLAS Phase-II Resistive Plate Chamber (RPC) detectors~\cite{Bauer:2019vqk}. This well-established technology requires no major R$\&$D, offering a reliable, cost-effective, and high-performance solution for tracking LLPs in ANUBIS.
The tracking stations will be arranged into two main layers separated by 1~m, each of which contains a triplet of RPCs stacked on top of each other. 
A potential single additional RPC layer between the two main layers is being studied in terms of its potential benefit for pattern recognition.
The RPC modules with approximate dimensions of $\sim 1\times2~\metre^2$ are tiled to fully cover the ceiling with a 5 cm overlap to ensure full acceptance, see Figure~\ref{fig:ANUBIS_Geometry}~(right). 
This would require $\mathcal{O}(6,000)$ RPC modules.
The active LLP decay volume of ANUBIS is then treated as originating from $\sim$2~m after the ATLAS calorimeters within the muon spectrometer, and extends up to 30~cm below the first ANUBIS tracking station.

The ANUBIS detector concept hinges on a full integration with the ATLAS experiment, thereby dramatically extending ANUBIS' sensitivity. 
For example, if a suitable displaced vertex candidate is reconstructed in its active decay volume, ANUBIS will have the capacity to trigger the ATLAS detector to obtain a full picture of the $pp$ collision event where the LLP was produced.
Similarly, ATLAS will have the ability to trigger ANUBIS. 
Thus, a set of LLP models that are produced in association with prompt objects can be explored, which is not possible for all LLP experiments. 
The ATLAS detector can also be used to strongly reduce the potential backgrounds of ANUBIS:
\begin{itemize}
\item 
ATLAS can act as a {\em passive} veto against SM LLPs, $K_L$ and neutrons. 
The calorimeter alone has an attenuation depth of $9.7~\Lambda_I$ for $|\eta|=0$, where $\Lambda_I$ is the nuclear interaction length.
Accounting for support structures, the solenoid magnet, and upstream detectors further increases the attenuation length, especially for $|\eta|>0$.
Moreover, the geometry of the ATLAS experiment has been well mapped allowing ANUBIS to veto against displaced vertices originating from within the ATLAS detector material and hence likely hadronic interactions. 
\item 
ATLAS can be used as an {\em active} veto against  high energy charged tracks and `punch-through jets' (jets that are not fully contained in the calorimeter) that are likely to produce displaced vertex candidates from hadronic interactions. 
Hence, displaced vertices aligned with punch-through jets and energetic charged tracks are vetoed. 
\end{itemize}

Overall, the air-filled active decay volume of ANUBIS combined with the symbiosis between ANUBIS and ATLAS is expected to allow ANUBIS to operate in a very low background scenario despite its relatively close proximity to the IP. 
This has been validated by data-driven estimates of the background using ATLAS analyses of displaced vertices in the ATLAS muon spectrometer~\cite{ATLAS:2018tup,ATLAS:2025pak}, and rescaling those to the full ANUBIS detector. 
From this we expect at most $\sim75$ background events at the HL-LHC assuming an integrated luminosity of 3~ab$^{-1}$.

\subsection{Sensitivity Studies of the ANUBIS Detector}
The potential sensitivity of the ANUBIS detector to several LLP models has been estimated at the HL-LHC assuming $\mathcal{L}=3\text{ ab}^{-1}$. 
The models selected were based on a set of recommended benchmarks outlined by the CERN Physics Beyond Colliders~(PBC) study group~\cite{Alemany:2019vsk}. 
A particular focus was placed on BC5, which involves an exotic Higgs boson decay to a pair of scalar LLPs, which in turn can mix with the Higgs boson in order to decay to SM fermions; and BC6, BC7, and BC8, which involve the introduction of a massive right-handed neutrino known as a Heavy Neutral Lepton (HNL) that couples only to the $\nu_e$, $\nu_\mu$, and $\nu_\tau$, respectively. 
In these studies, two scenarios were examined: a zero background case, corresponding to an observation with four LLP signal events detected; and a conservative background case, corresponding to an observation with 90 LLP signal events. 
The latter is based on a conservative upper bound from a data-driven background estimate from measurements in Ref~\cite{ATLAS:2018tup} summarised above.

\begin{figure}[tb]
    \centering
    \includegraphics[width=0.35\linewidth]{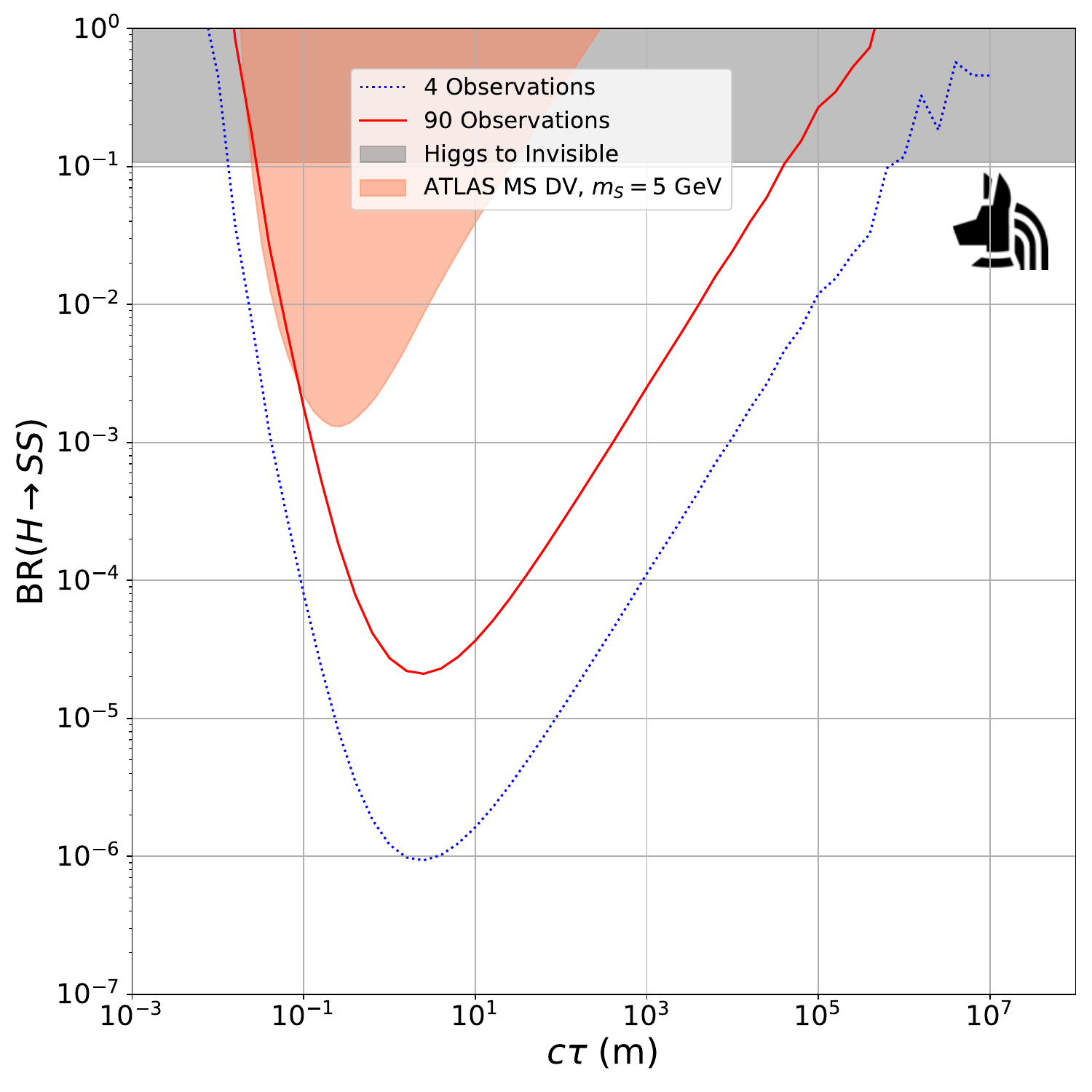}
    \includegraphics[width=0.35\linewidth]{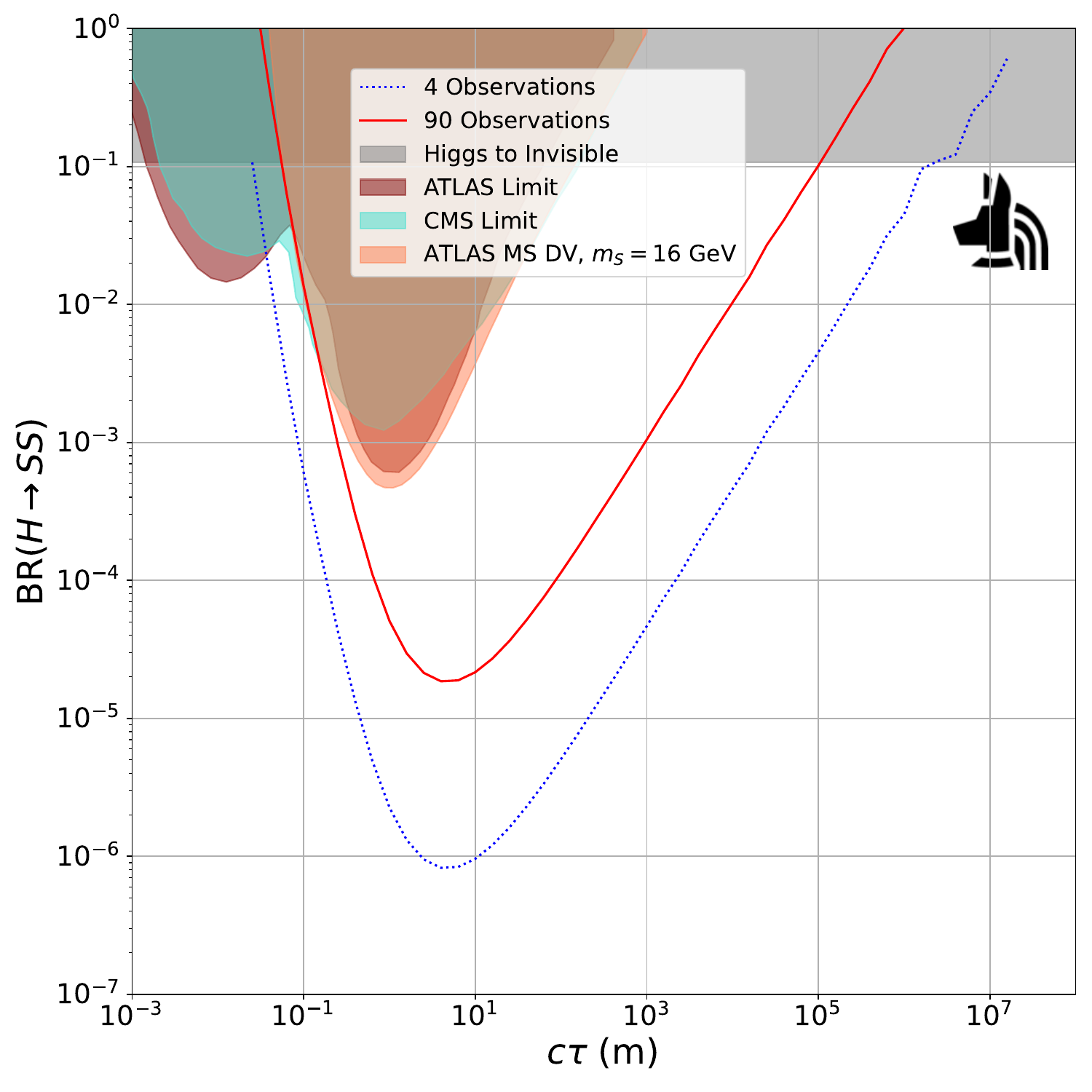}
    \includegraphics[width=0.35\linewidth]{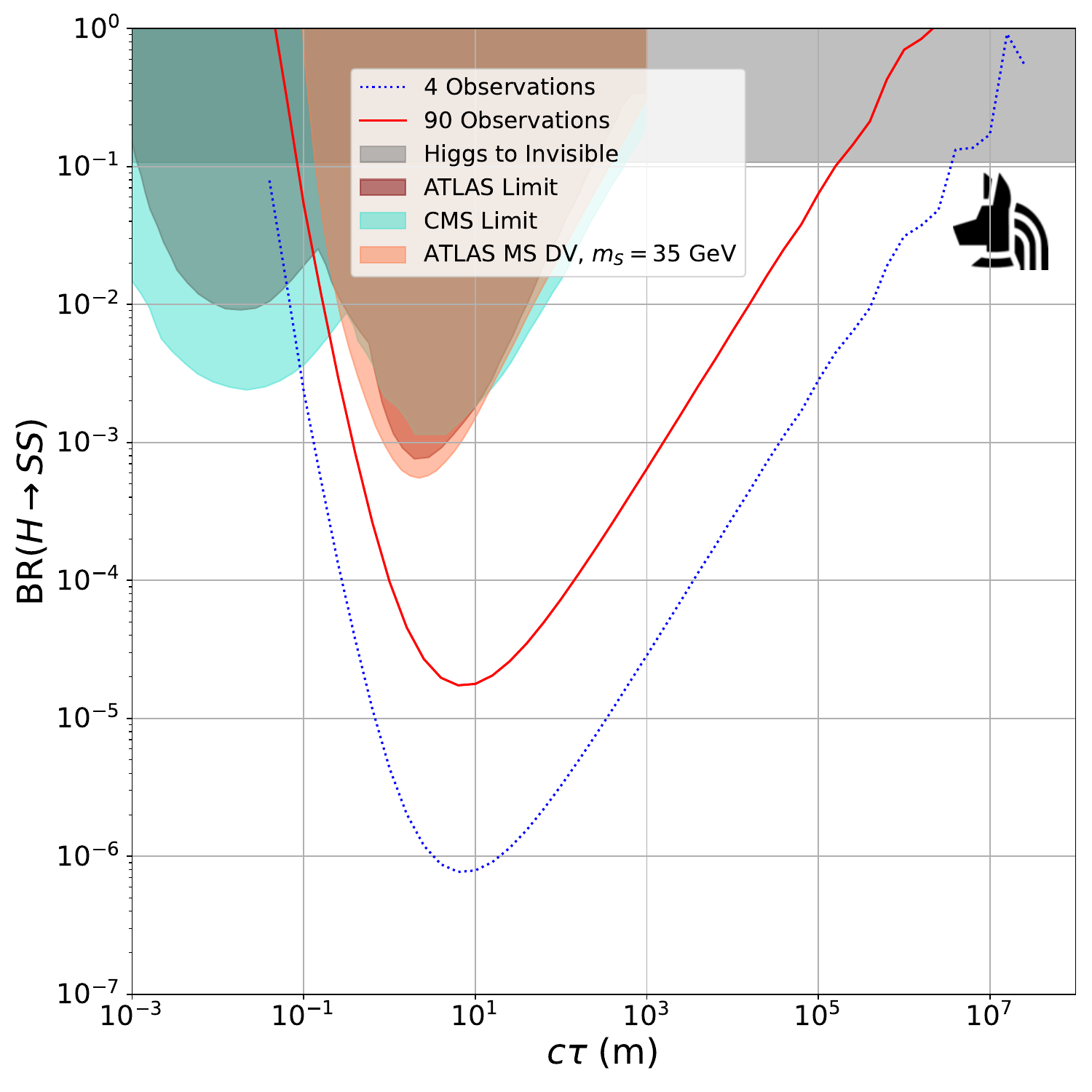}
    \includegraphics[width=0.35\linewidth]{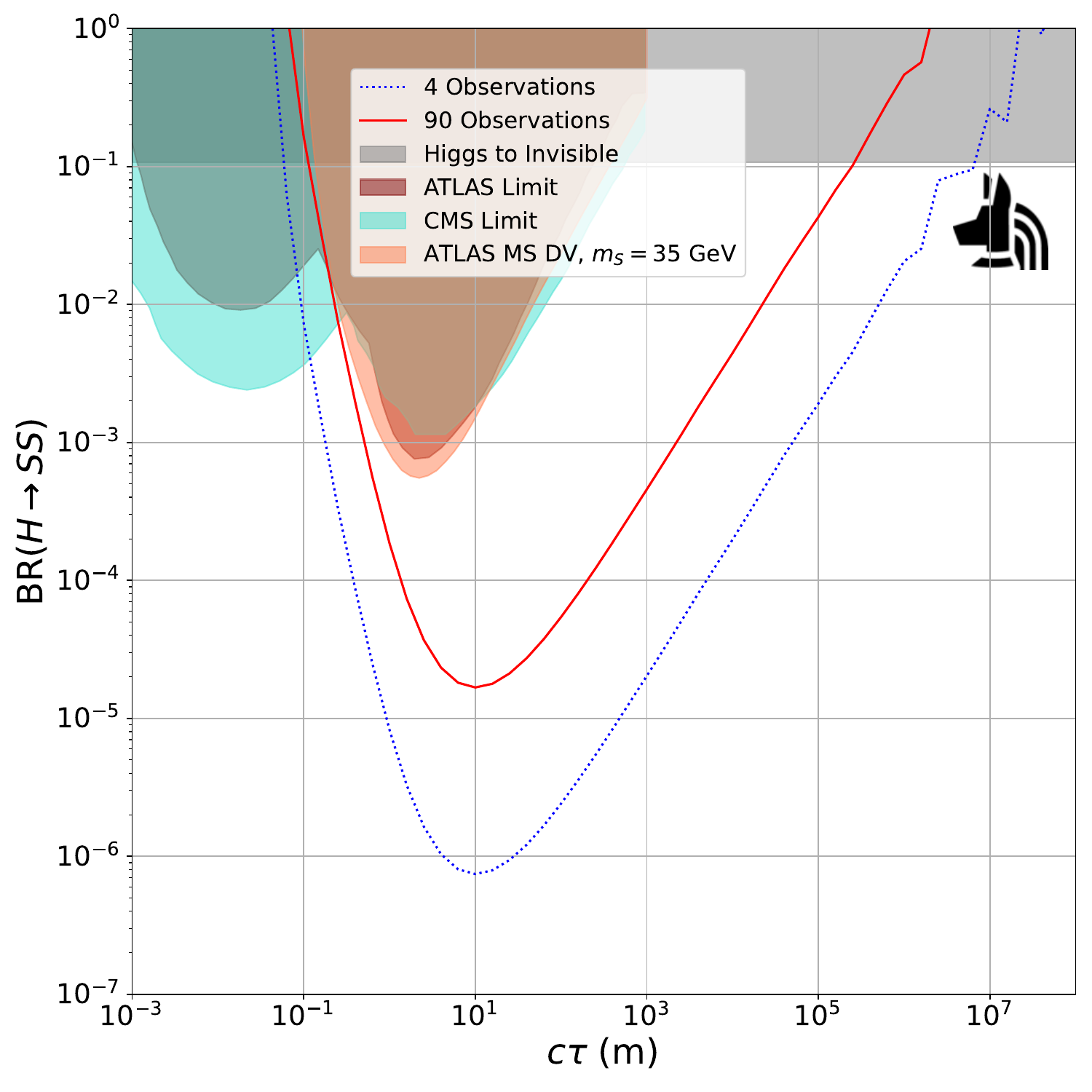}
    \caption{Projected sensitivity of the ANUBIS detector to $H\rightarrow SS$, where $S\rightarrow b\bar{b}$. The limits are shown for the case of zero expected background and a conservative background requiring detecting four and 90 signal events for observation. The Higgs to invisible limit~\cite{ATLAS:2023tkt}, as well as the combined limits for ATLAS~\cite{ATLAS:2018pvw,ATLAS:2024qoo,ATLAS:2024ocv,ATLAS:2022gbw,ATLAS:2023tkt,ATL-PHYS-PUB-2021-020} and CMS~\cite{CMS:2024zfv,CMS:2024xzb,CMS:2024bvl,CMS:2023sdw} are overlaid. The results of an ATLAS LLP search using a single displaced vertex in the muon spectrometer~\cite{ATLAS:2025pak} are shown separately.}
    \label{fig:HtoSS_Sensitivity}
\end{figure}

The BC5 model was chosen to highlight ANUBIS' ability to target heavy mediators at the electroweak scale. Figure~\ref{fig:HtoSS_Sensitivity} shows the lower limit of the branching ratio of $H\rightarrow SS$, where $S\rightarrow b\bar{b}$, over a range of $c\tau$ for scalar LLP masses of 10, 20, 30 and 40 GeV~\cite{Satterthwaite:2839063}. The maximal sensitivity is around a $c\tau$ value of 10 m, where the expected sensitivity can reach down to $\mathcal{B}(H\rightarrow SS)\lesssim 10^{-6}$, four orders of magnitude below the limits from searches for invisible Higgs decays. 
Additionally, ANUBIS demonstrates an improvement of two to three orders of magnitude compared to the latest ATLAS results~\cite{ATLAS:2018pvw,ATLAS:2024qoo,ATLAS:2024ocv,ATLAS:2022gbw,ATLAS:2023tkt,ATL-PHYS-PUB-2021-020}. 

\begin{figure}[tb]
    \centering
    \includegraphics[width=0.48\linewidth]{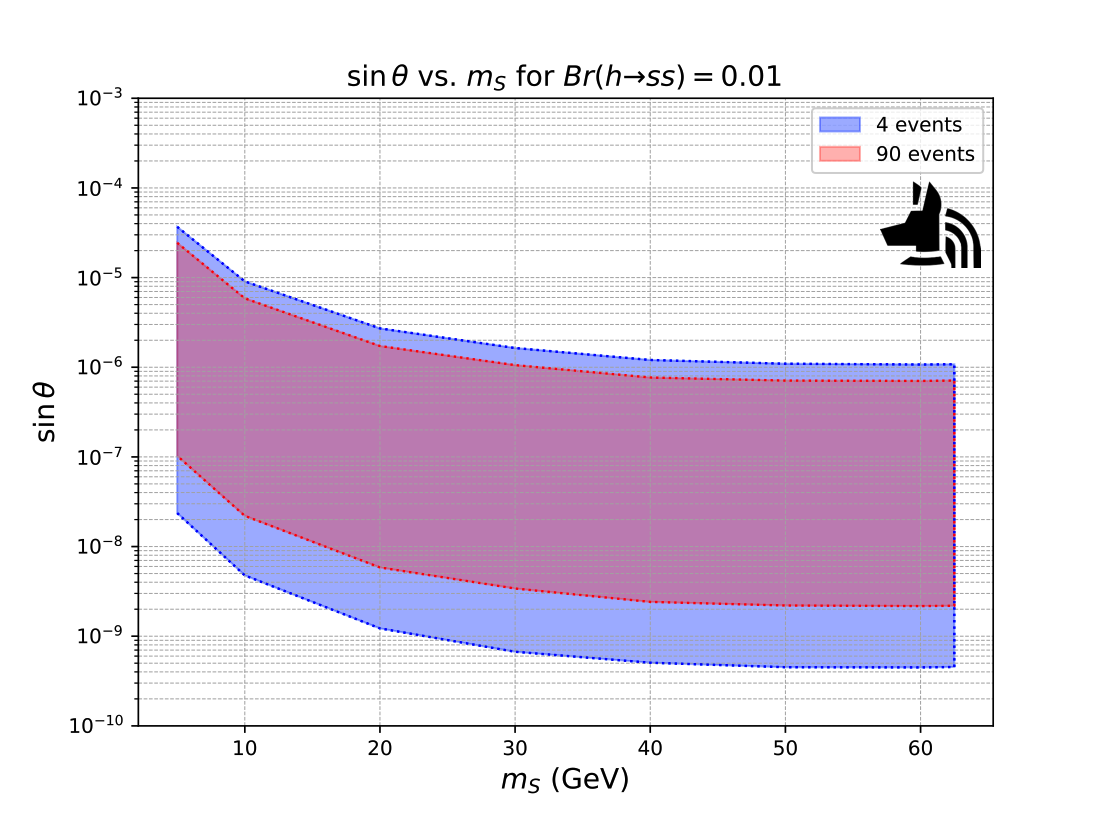}
    \includegraphics[width=0.48\linewidth]{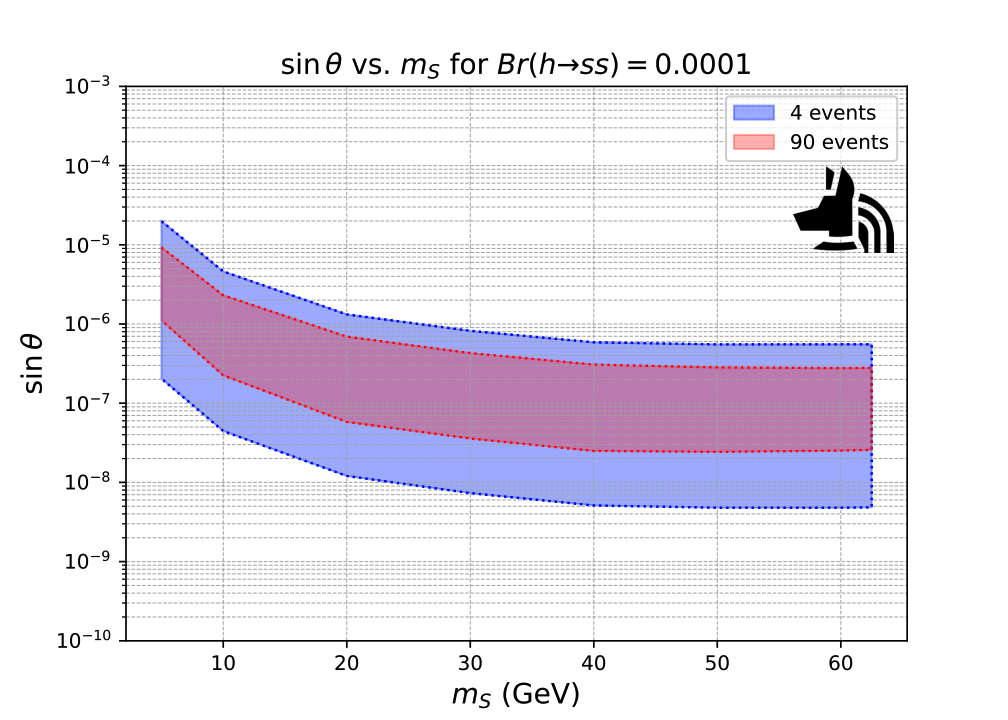}
    \caption{Projected sensitivity of the ANUBIS detector to $H\rightarrow SS$, where $S\rightarrow b\bar{b}$, in terms of the mixing angle, $\theta$, for (left) $\mathcal{B}(H\rightarrow SS)=0.01$ and (right) $\mathcal{B}(H\rightarrow SS)=10^{-4}$.}
    \label{fig:HtoSS_Sensitivity_BC5}
\end{figure}

In BC5, the key parameter of interest is \(\sin\theta\), where \(\theta\) represents the mixing angle between the BSM scalar and the Higgs boson, assuming a fixed branching ratio \(\mathcal{B}(H \to SS)\). The limits on \(\mathcal{B}(H \to SS)\) as a function of \(c\tau\) (Figure~\ref{fig:HtoSS_Sensitivity}) have been translated into constraints on \(\theta\) assuming \(\mathcal{B}(H \to SS) = 0.01\), as shown in Figure~\ref{fig:HtoSS_Sensitivity_BC5} (left). The choice of \(\mathcal{B}(H \to SS) = 0.01\) follows the BC5 PBC benchmark. However, since ANUBIS has the capability to probe significantly lower branching ratios, this benchmark may underestimate its full sensitivity. To better capture ANUBIS' reach, the limits are also evaluated for \(\mathcal{B}(H \to SS) = 10^{-4}\) and presented in Figure~\ref{fig:HtoSS_Sensitivity_BC5} (right). Remarkably, despite the branching ratio being reduced by two orders of magnitude, the lower bound on \(\sin\theta\) remains within approximately one order of magnitude. At such small branching ratios only transverse experiments have any sensitivity for $c\tau\gtrsim10~$m.

\begin{figure}
    \centering
    \includegraphics[width=0.65\linewidth]{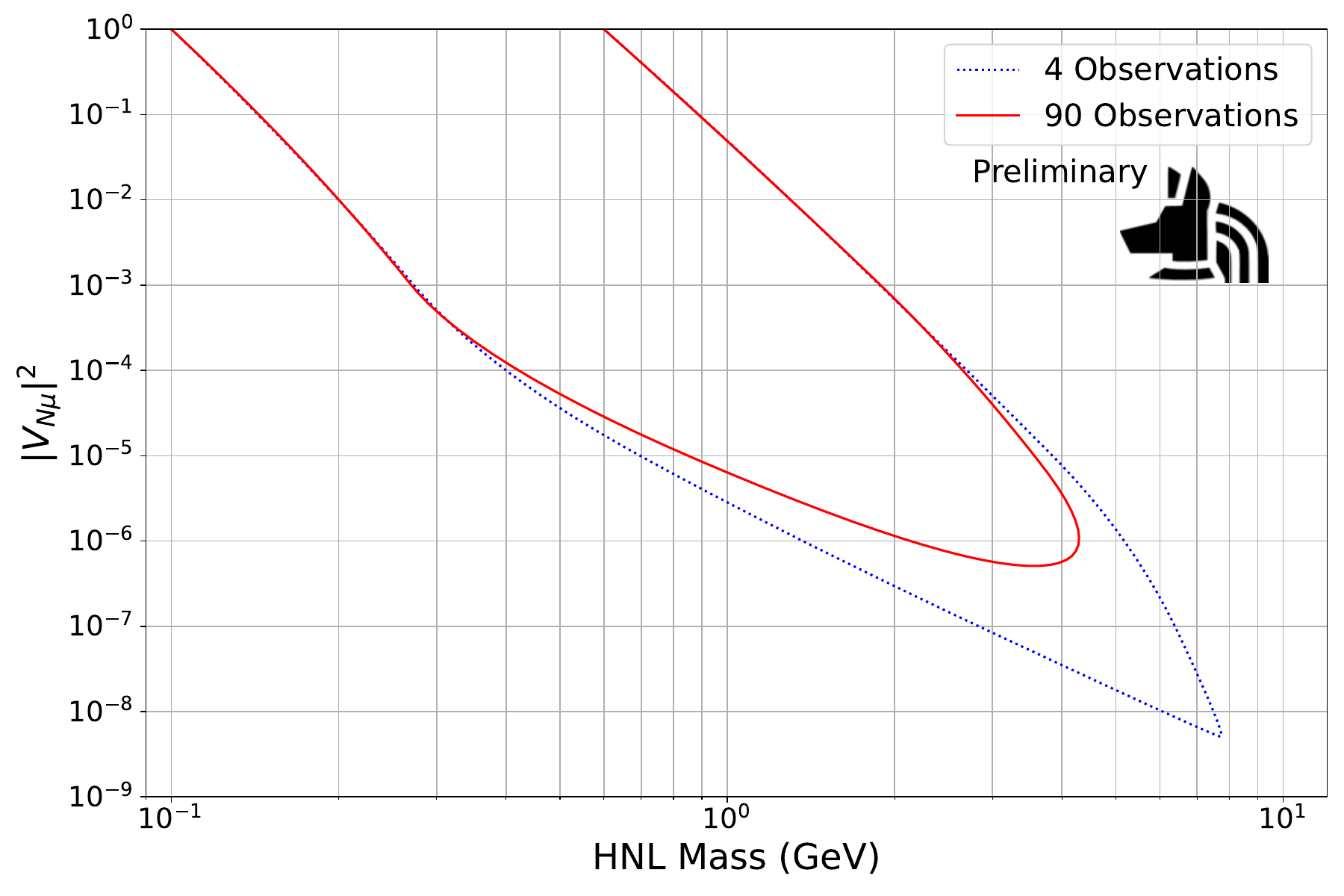}
    \caption{Preliminary projected sensitivity of ANUBIS to the muon-coupled HNL (BC 7) in terms of the muon-HNL coupling $|V_{N\mu}|^2$ and the HNL mass in the case of zero expected background (4 observations, blue) and for moderate expected background (90 observations, red).}
    \label{fig:HNL_Sensitivity}
\end{figure}

For BC6, BC7, and BC8, a single right-handed Majorana neutrino is introduced via a Type I Seesaw mechanism~\cite{PhysRevD.100.075029}. This model highlights the ability of ANUBIS to probe LLPs directly at the electroweak scale through Charged-Current and Neutral-Current Drell-Yan production of the HNLs. 
To estimate ANUBIS' sensitivity, the HNL decays into charged final states ${N\rightarrow l^{\pm}qq^\prime}$,  $N\rightarrow \nu_lqq^\prime$, $N\rightarrow l_i^+l_i^-\nu_{l_j}$ that can be observed at ANUBIS have been considered.
The electroweak production modes and the visible decay modes are integrated together to determine the sensitivity of ANUBIS to these models. Figure~\ref{fig:HNL_Sensitivity} shows the result from a preliminary study focusing on the muon-coupled HNL (BC7) to provide an approximate first sensitivity projection. 
This shows that ANUBIS can probe very low values of the muon--HNL coupling for HNL masses $\mathcal{O}(1-10)\text{ GeV}$, and would provide excellent sensitivity around the scale of $m_W$ and $m_Z$.

\subsection{The proANUBIS Detector}
\begin{figure}[htpb]
    \centering
    \includegraphics[width=0.48\linewidth]{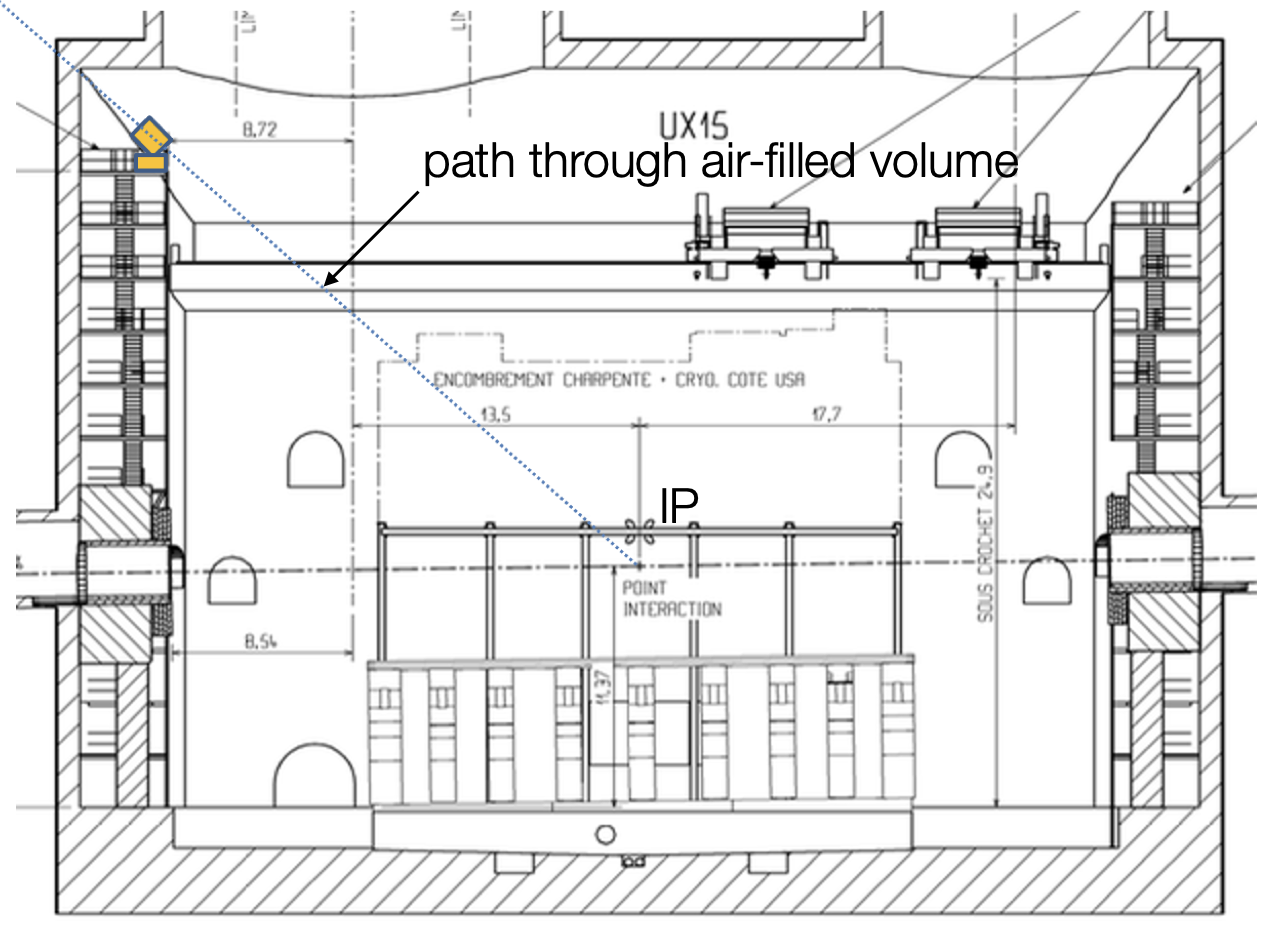}
    \includegraphics[width=0.3\linewidth]{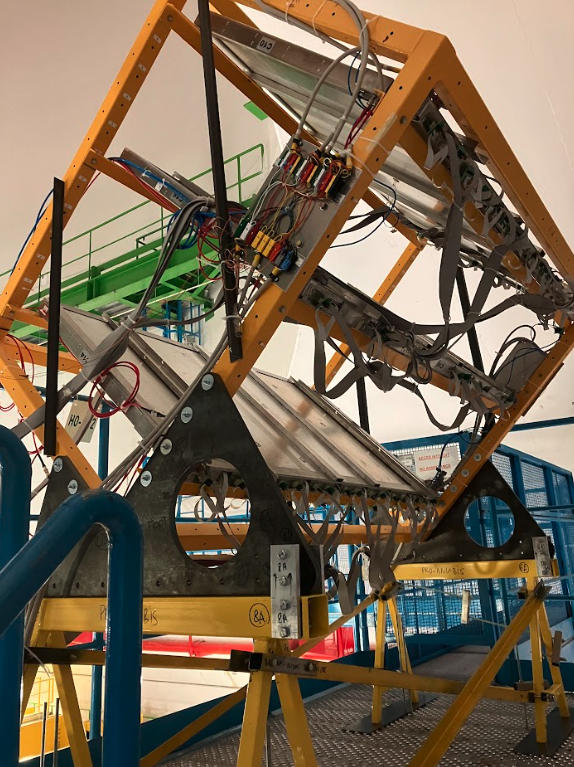}
    \caption{(left) The ATLAS Experimental Cavern with the proANUBIS detector (yellow) overlaid in its position. (right) The proANUBIS detector installed in position.}
    \label{fig:proANUBIS_Detector}
\end{figure}

A prototype of ANUBIS, known as proANUBIS, was installed in the ATLAS cavern and has been collecting data since March 2024. To date, it has recorded $104\text{ fb}^{-1}$ of $pp$ collision data.
The proANUBIS demonstrator was placed at the top of a support structure within the ATLAS cavern close to the ceiling, the nominal location of ANUBIS, as shown in Figure~\ref{fig:proANUBIS_Detector}~(left). The detector is inclined at an angle such that its tracking layers are perpendicular to the IP~({\em ibid},~right). 
The proANUBIS structure consists of three layers of RPC modules in a similar geometry as a single tile of ANUBIS. However, instead of a triplet-singlet-triplet design, proANUBIS adopts a doublet-singlet-triplet design, where the doublet is furthest from the IP. 

The primary physics goal of the proANUBIS demonstrator is to provide a detailed measurement of the expected backgrounds {\em in situ}. Additionally, it provides valuable insights into the performance of the chosen detector technology and its operation in an environment very similar to the full ANUBIS design.

\begin{figure}[htpb]
    \centering
    \includegraphics[width=0.8\linewidth]{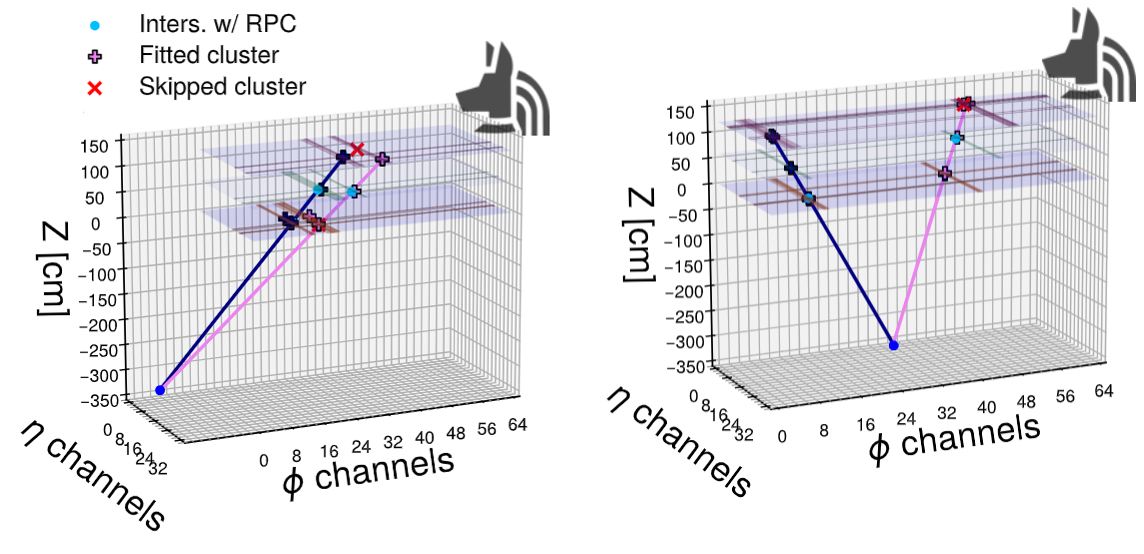}
    \caption{
    Two reconstructed vertices from proANUBIS $pp$ data, where hits on the RPC planes are clustered and fitted to determine the vertex location.}
    \label{fig:proANUBIS_Vertices}
\end{figure}

The initial analysis of the 2024 dataset has significantly advanced our track and vertex reconstruction capabilities, enabling the effective reconstruction of two- and three-track vertices, with a couple of representative examples shown in Figure~\ref{fig:proANUBIS_Vertices}. 
Moreover, proANUBIS has been synchronised against the LHC clock provided by the Central Trigger Processor (CTP) of ATLAS. 
This synchronisation is required as proANUBIS is not directly integrated into the ATLAS trigger system like the ANUBIS would be. 
Instead, a copy of the ATLAS Bunch Crossing Reset and Level-1 triggers is recorded alongside proANUBIS data.  
By leveraging ATLAS data, we observe a global bunch crossing offset of approximately $\sim114$ between ATLAS and proANUBIS, corresponding to a latency of 2850 ns. This offset is dominated by a combination of electronic delays in the CTP system, with smaller contributions from signal propagation and time of flight. 
The proANUBIS demonstrator was shown to be able to select muons in ATLAS data within an angular proximity of $\Delta R=\sqrt{\phi^2 + \eta^2}=0.2$ to the proANUBIS detector, specifically for events with time stamps that are consistent with those of proANUBIS. This allows the identification of candidate muons in ATLAS data that were simultaneously recorded by proANUBIS.  

Ultimately, this synchronisation capability demonstrates the feasibility of utilising ATLAS prompt information to veto background events in the full ANUBIS detector. Moreover, it also allows for a direct study of the efficiency of such a veto using proANUBIS. This, after full analysis, will help to validate the feasibility of implementing an active veto strategy for the full ANUBIS detector, leveraging ATLAS data to enhance its background rejection capabilities.


\section{Readiness and expected challenges}
The readiness and expected challenges to realise ANUBIS can be summarised as follows:
\begin{itemize}
\item 
The RPC technology is mature. A prototype demonstrator proANUBIS has collected 104~\fb of $pp$ collision data using the ATLAS Phase I RPC technology. 
Overall, an excellent performance with only a year of data taking has been demonstrated, providing a clear proof of concept.
\item 
The ANUBIS detector technology will be based on the ATLAS Phase II RPC technology, which is very similar to Phase I and comes with additional significant cost savings by integrating a Time-to-Digital Converter directly into the front-end on-detector electronics.
\item 
A minor challenge is that the Data Acquisition (DAQ) requires some R\&D for potential large cost savings by synchronising (daisy-chaining) the readout.
\item 
Another minor challenge is firming up mechanical design with LHC engineering team. In initial discussions, no showstoppers have been identified.
\item 
An additional challenge is that the current gas mixture employed in the RPC chambers has a high global-warming potential. Therefore, a more environmentally friendly gas mixture needs to be identified and employed for the operation of the full ANUBIS detector. Significant progress has been made, and promising alternative gas mixtures are being tested.
\item 
One of the main challenges is finding a viable storage solution during LHC shutdowns for the tracking stations at the bottom of the PX14 and PX16 service shafts.
\item 
A straightforward challenge to be addressed is the full integration of DAQ and Detector Control System (DCS) with ATLAS.
\end{itemize}

\section{Timeline}
\begin{itemize}
    \item \textbf{2019}: ANUBIS was proposed as a conceptual detector~\cite{Bauer:2019vqk}.
    \item \textbf{2020}: An initial sensitivity study was performed using PBC Benchmark 5 (BC 5).
    \item \textbf{2021}: The proANUBIS detector was conceived as a prototype demonstrator to be located in the ATLAS cavern.
    \item \textbf{2022}: Seed funding (1.5M CHF) obtained for ANUBIS and proANUBIS. 
    \item \textbf{2023}: The proANUBIS demonstrator was finalised, constructed, deployed, and commissioned in the ATLAS Cavern with support of CERN and ATLAS technical coordination. In the same year, proANUBIS was incorporated as a sub-project of ATLAS.
    \item \textbf{2024}: 104~\fb of $pp$ collision data recorded by proANUBIS. Simultaneous work on PBC benchmarks 7, 8 and 9 started, as well as a full \geant{4} model for ANUBIS. Additionally, contributed to the ATLAS Phase II RPC upgrade.
    \item \textbf{2025}: Ongoing data analysis of the proANUBIS dataset. Letter of Intent for the full ANUBIS detector to be submitted.
    \item \textbf{2026+}: ANUBIS detector R\&D including the readout electronics and engineering for the full installation in the ATLAS Cavern.
    \item \textbf{2028+}: Partial deployment of ANUBIS during Long Shutdown 3 (LS3) in the LHC schedule. Full integration with the ATLAS DAQ and DCS systems.
    \item \textbf{2030+}: Data taking with the partially deployed ANUBIS detector in LHC Run 4.
    \item \textbf{2033+}: Full deployment of the ANUBIS detector during LS4.
    \item \textbf{2035+}: Data taking with the full ATLAS+ANUBIS detector during LHC Run 5+. 
\end{itemize}

\section{Construction and operational costs}
The main cost challenge for ANBUBIS comes from its large instrumented area of about 1,600~$\metre^2$. 
This is overcome by using RPC technology which fulfills all detector requirements for a moderate price tag of below 10~kCHF/$\metre^2$ including on-detector electronics.

The core construction costs for ANUBIS are expected to fall within the range of 20 to 30 MCHF, including R\&D, civil engineering, construction, and installation tooling. Given this moderate price tag, ANUBIS is our preferred TPF to fully exploit the physics potential of the HL-LHC by extending its sensitivity to DM and baryogenesis models with LLP signatures by orders of magnitude.

There are only minor civil engineering costs associated with ANUBIS since the concept hinges on using existing infrastructure. The main contribution to the core costs comes from the detectors themselves (gas gap volumes, pickup strips, front-end electronics) as well as the DAQ system (serialisers, data transmission links, trigger and timing system), which account for about 20 MCHF.

\subsubsection*{Operation costs}
The operation costs of ANUBIS are fairly low, as they only involve the operation and maintenance costs of the gas system (200~kCHF/year), electrical power provision~(50~kCHF/year) and maintenance, as well as the maintenance of the DAQ system including spares and minor consumables (50~kCHF/year). 
The computing for the triggering and DAQ is another cost factor that contributes about 200~kCHF/year, resulting in a total of about 500~kCHF/year during data taking, not accounting for inflation.

The above does not account for the time of researchers needed to operate ANUBIS. 
However, such costs will be kept low by fully integrating ANUBIS into the ATLAS DCS system, which would allow combining it with the regular ATLAS muon system operations.

\clearpage 
\printbibliography[title=References,heading=bibintoc]

\end{document}

%% file: style.tex
\def\metre{\text{m}\xspace}

\def\mllp{\ensuremath{m_\text{LLP}}\xspace}
\def\taullp{\ensuremath{\tau_{\rm LLP}}\xspace}

\newcommand{\nc}{\newcommand}
\nc{\beq}{\begin{equation}}
\nc{\eeq}{\end{equation}}
\nc{\barray}{\begin{eqnarray}}
\nc{\earray}{\end{eqnarray}}
\nc{\barrayn}{\begin{eqnarray*}}
\nc{\earrayn}{\end{eqnarray*}}
\nc{\bcenter}{\begin{center}}
\nc{\ecenter}{\end{center}}
\nc{\mc}{\mathcal}
\newcommand{\gsim}{\lower.7ex\hbox{$\;\stackrel{\textstyle>}{\sim}\;$}}
\newcommand{\lsim}{\lower.7ex\hbox{$\;\stackrel{\textstyle<}{\sim}\;$}}

\newcommand{\etmiss}{\ensuremath{E \kern-0.6em\slash_{\rm T}}\xspace}
\newcommand{\etmissx}{\ensuremath{E \kern-0.6em\slash_{\rm x}}\xspace}
\newcommand{\etmissy}{\ensuremath{E \kern-0.6em\slash_{\rm y}}\xspace}

\newcommand{\geant}{{\sc Geant}\xspace}

\newcommand{\GeV}{\ensuremath{\textnormal{GeV}}\xspace}

\newcommand{\fb}{\ensuremath{{\rm fb}^{-1}}\xspace}

\definecolor{myorange}{rgb}{0.99,0.88,0.8}

\newtcolorbox{mybox}{
    arc=0pt,
    boxrule=0pt,
    colback=myorange,
    width=1.\textwidth,   
    colupper=black,
}